\documentclass[a4paper,fleqn,usenatbib,useAMS]{mnras}

\usepackage{graphicx}	
\usepackage{amsmath}	
\usepackage{amssymb}	
\usepackage{multicol}	
\usepackage{bm}			
\usepackage{pdflscape}	
\usepackage[T1]{fontenc}
\usepackage{ae,aecompl}
\usepackage{txfonts}

\title[A Second Decoupling During Black-hole Merger]{A Second Decoupling Between Merging Binary Black Holes
and the Inner Disc--Impact on the Electromagnetic Counterpart} 
\author[C. Fontecilla, X. Chen \& J. Cuadra]{Camilo Fontecilla$^{\dagger}$,
Xian Chen$^{\dagger,}$\thanks{E-mail: xchen@astro.puc.cl}
and Jorge Cuadra$^{\dagger}$
\\
$^{\dagger}$ Instituto de Astrof\'isica, Pontificia Universidad Cat\'olica de Chile, Av. Vicu\~na Mackenna 4860, Santiago, Chile}

\date{Accepted 20XX . Received 20XX}
\pubyear{2016}

\begin{document}
\label{firstpage}
\pagerange{\pageref{firstpage}--\pageref{lastpage}} 
\maketitle

\begin{abstract}
The coalescence of two supermassive black holes (SMBHs) produces  powerful
gravitational-wave (GW) radiation and, if gas is present in the vicinity, also
an electromagnetic (EM) counterpart. In the standard picture, an EM outburst
will be produced when the binary ``decouples'' from the circum-binary disc and
starts ``squeezing'' the disc inside the secondary orbit, resulting in its
quick accretion on to the primary black hole.  Here we use analytical arguments
and numerical simulations to show that the disc within about $20~R_S$ of a SMBH
survives the merger without being depleted. The reason is a ``second
decoupling": the inner disc thickens due to tidal heating and inefficient
cooling, effectively decoupling from the interaction of the binary. We show
that this second decoupling quenches the heating sources in the disc ${\cal
O}(10^2)$ days before coalescence.  This will render the peak UV/X-ray
luminosity significantly weaker than previously thought. After the merger, the
residual disc cools down and expands, merging with the outer disc rather than
being completely accreted. This results in  continuous EM emission, hindering
the detection of the cut-off and re-brightening proposed in earlier studies.
\end{abstract}


\section{Introduction}
\label{sec:intro}

Supermassive black hole binaries (SMBHBs)  form in galactic nuclei following
hierarchical galaxy mergers \citep{begelman80,volonteri03}.  Depending on the
conditions of the surrounding stellar and gaseous environment, some SMBHBs
efficiently lose orbital energy and angular momentum, managing to coalesce
within a Hubble time \citep[see][ for reviews]{merritt05,colpi09}.  These
mergers produce powerful gravitational wave (GW) radiation whose frequency
falls in the sensitive bands of the ongoing Pulsar Timing Array
project\footnote{http://www.ipta4gw.org} and the planned Laser Interferometer
Space Antenna\footnote{https://www.elisascience.org}. 

Electromagnetic (EM) radiation could be produced simultaneously with GW
emission if a merger happens in an gas-rich environment \citep[see][for
reviews]{centrella10,schnittman13}. Such ``EM counterparts'' contain rich
information about the distribution and evolution of matter around supermassive
black holes (SMBHs).  This information, combined with those extracted from GW
signals, would deliver a more comprehensive picture in the strong-gravity
regime.  Such a picture will greatly enrich our knowledge of black hole physics
and astrophysics.

A particular place that can produce EM counterparts is active galactic nucleus
(AGN). SMBHBs in AGNs are likely embedded in gaseous accretion discs.
Theoretical models predict that after a binary shrinks to a separation of
$a\sim10^2$ Schwarzschild radius ($R_S$) it will ``decouple'' from an ``outer
disc''--the part of the disc outside the binary orbit-- because the merger
time-scale of the SMBHs due to GW radiation becomes shorter than the viscous
time-scale of that disc \citep{armitage02}. More importantly for this work,
since the GW radiation time-scale diminishes quickly as the binary separation
shrinks, it will become shorter than the viscous time-scale of the ``inner
disc''--the disc enclosed by the binary orbit. The consequence is a strong
tidal ``squeezing'' of the inner disc by the coalescing binary
\citep{armitage02}.

It is generally accepted that the squeezing would heat up the inner disc and
produce a luminous ``precursor''--an enhancement of the EM radiation-- days to
hours before the   SMBHB merger \citep{lodato09,chang10,tazzari15,cerioli16}.
It is also suggested that this process could drive the material of the inner
disc either inward into the SMBHs or outside the binary orbit, so that
eventually there is no material near the post-merger SMBH \citep{armitage02}.
As a result, AGN activity is halted \citep[such as jet
formation][]{liu03,liu04} and no UV/X-ray radiation could be detected until the
outer disc either refills the cavity
\citep{milosavljevi05,dotti06,chang10,tanaka10,shapiro10} or gets shock-heated
either by the sudden change in the gravitational potential
\citep{schnittman08,megevand09,oneill09} or the recoil of the central SMBH
\citep{lippai08,shields08,schnittman08,rossi10}.
 
However, an important point has been often overlooked in the previous studies:
the squeezing mechanism is effective only when the inner disc is geometrically
thin. If the disc becomes thick, fluid elements can cross the binary orbit and
leak to the outer disc, either due to a large effective viscosity
\citep{lin79,papaloizou84} or through horseshoe trajectories
\citep{baruteau12}. \citet{armitage02} envisaged that the inner disc inevitably
becomes geometrically thick based on the observation that in their numerical
simulations when $a\la10~R_S$ the accretion rate exceeded the Eddington limit.
\citet{tazzari15} also mentioned the effect of tidal heating on the thickness
of the disc, but they did not include it in their models.

In this Letter we use both analytical arguments and numerical simulations to
prove that at a relatively large binary separation, $a\ga20~R_S$, the inner
disc always becomes geometrically thick. We also discuss the impact on the
detectability of the EM counterparts.

\section{Heating sources and disc thickness}\label{sec:heat}

The thickness of an accretion disc can be characterized by the scale height $h$
(or ``half thickness'') and is normally supported by gas pressure $p_{\rm gas}$
and radiation pressure $p_{\rm rad}$. We are interested in the case when
radiation pressure dominates since we are considering a disc that is at a
distance of $r\la10^2R_S$ from the SMBH. In this case, the condition for
vertical hydrostatic equilibrium reduces to 
\begin{equation} 
h=\kappa F/(c\Omega^2),
\label{eq:F}
\end{equation}
where $F$ is the radiation flux at the disc surface, $\Omega$ the Keplerian
angular velocity, $c$ the speed of light and $\kappa$ the opacity. 

To quantify $h$, we first consider a standard thin disc where radiative
cooling is balanced by viscous heating. This condition leads to the equation of
energy equilibrium \citep{frank02}
\begin{equation}
F = D_\nu = 9 \nu \Sigma \Omega^2 / 8
\label{eq:Dnu},
\end{equation}
where $D_\nu$ is the viscous dissipation rate per unit surface area, $\nu$ the
viscosity and $\Sigma$ the surface density of the disc. Assuming a stationary
disc, we can substitute $\nu \Sigma$ with a constant fraction of the mass
accretion rate, $\dot{M}/ (3 \pi)$ \citep{frank02}.  From Equations
(\ref{eq:F}) and (\ref{eq:Dnu}) we find that the scale height has a constant
value $h_0 = 3 \kappa \dot{M} / (8 \pi c) \simeq 7.4 \dot{m} R_S$, depending
only on $\dot{m}$, the accretion rate normalized by the Eddington rate
$\dot{M}_{\rm Edd}$, assuming a mass-to-radiation coefficient of $0.1$. 

The above analysis indicates that when the accretion rate of a disc approaches
the Eddington limit, i.e. $\dot{m}\sim1$, the inner part of it, e.g.
$r\la10R_S$, will have an aspect ratio of $h_0/r\ga1$ and the disc will become
thick.  The same conclusion applies to the squeezing phase, only that the
accretion rate and the heating rate will be determined by the tidal
interaction, not viscosity.

The squeezing phase starts when the GW radiation time-scale of a SMBHB, $t_{\rm
GW}(a)$, becomes shorter than the viscous time-scale of the inner disc, $t_{\rm
vis}$. For a circular binary,
\begin{equation}
t_{\rm GW}(a) = 5 a^4 / \left[8 c R_S^3 q (1 + q) \right]
\label{eq:tgw}
\end{equation}
\citep{peters64}, where $q=M_s/M_p$ is the mass ratio of the binary, $M_p$ is
the mass of the primary (bigger) black hole, $M_s$ is the mass of the secondary
one.  For simplicity, we only consider the disc surrounding the primary SMBH,
but our conclusions can also be applied to the disc around the secondary.

We can, without loss of generality, consider the fluid elements of a thin
annulus between the radii $r$ and $r + \Delta r$, where $\Delta r\ll r$ is the
width of the annulus. When $t_{\rm GW}\ll t_{\rm vis}$,  the squeezing
mechanism will force the radius of this ring to shrink on a time-scale similar
to the shrinking time-scale of the SMBHB. Therefore, if $v_r$ is the radial
velocity of the annulus, we have $|r / v_r| \simeq t_{\rm GW}$.  To satisfy
this relationship, the fluid elements in the ring must lose their orbital
energy \citep[by shocks,][]{lin86b} at a rate of $(\pi v_r \Sigma \Delta r)(G
M_p / r)$, with $G$ the gravitational constant. This dissipation will heat up
the disc at a rate, per unit surface area, of
\begin{equation}
D_\Lambda \sim - G M_p \Sigma v_r / (4 r^2).
\label{eq:DL}
\end{equation}

This means that in this phase we have two heating terms in the energy equation: $D_\Lambda$ due to the tidal torque and $D_\nu$ due to viscosity.  Taking into account the fact that
$t_{\rm vis}= 2 r^2 / (3 \nu)$,  we can write
\begin{equation}
D_\Lambda / D_\nu \sim t_{\rm vis} / (3 t_{\rm GW}).
\end{equation}
This ratio is independent of the detailed structure of the disc and is valid as
long as the fluid elements of the inner disc do not cross the orbit of the
SMBHB.  Since we already know that $t_{\rm GW}(a) \ll t_{\rm vis}(r)$ during
the squeezing phase, it becomes clear that that $D_\Lambda\gg D_\nu$.

Therefore, the appropriate equation for thermal equilibrium is
$F=D_{\nu}+D_{\Lambda}\sim D_{\Lambda}$. 
From this we derive that
\begin{equation}
h_0 \simeq 2.5 \dot{m} R_S,\label{eq:h}
\end{equation}
where $\dot{m}=2\pi rv_r\Sigma/\dot{M}_{\rm Edd}$. Again we find that the inner
part of an accretion disc becomes thick when $\dot{m}\sim1$.

Although the dependence of $h$ on $\dot{m}$ is the same (linear) for both
viscosity- and tidally-dominated discs, the accretion rate $\dot{m}$ entails
very different physics in these two cases. For a standard disc, $\dot{m}$ is
determined by viscosity and hence proportional to $D_\nu$. In the case of a
squeezed disc, $\dot{m}$ is driven by the tidal force so it scales with
$D_\Lambda$.  If one mistakenly uses $D_\nu$ to calculate energy dissipation
during the squeezing phase, one would significantly underestimate the accretion
rate as well as the scale height, wrongly considering the disc to be thin.

For this reason, earlier works that neglected $D_\Lambda$
\citep{armitage02,baruteau12,tazzari15,cerioli16} inevitably have
underestimated $h$. \citet{lodato09} and \citet{chang10} included $D_\Lambda$
in their energy equations. However, they do not appear to have paid attention
to the aspect ratio, and hence overlooked that $h / r$ will be ${\cal O}(1)$
during the squeezing phase. In the following sections we will calculate $h/r$
and show that it gets close to unity for $a\sim20~R_S$.

\section{Analytical model}\label{sec:analytical}

To find out at which binary separation is the condition $h / r \simeq 1$ 
satisfied, we replace $v_r$ in Equation (\ref{eq:DL}) with $-r/t_{\rm GW}$,
where $t_{\rm GW}$ is a function of $a$. Then from Equation (\ref{eq:F}) and
$F=D_\Lambda$, we derive
\begin{equation} h/r= \sqrt{t_{\rm cool} / (2 t_{\rm GW})} \label{eq:hrtt},
\end{equation}
where $t_{\rm cool}=\tau h / c$ is the cooling time-scale due to radiation and
$\tau = \kappa \Sigma / 2$ the optical depth of the disc.
Equation~(\ref{eq:hrtt}) indicates that the thickness of the inner disc closely
correlates with its ability to cool. Only when $t_{\rm cool}\ll t_{\rm GW}$ is
the disc geometrically thin.

To proceed, we need to express $t_{\rm cool}$ as a function of $r$ as well.
This requires knowledge of $\Sigma$. Although we do not know yet the surface
density during the squeezing phase (this will be calculated in the next
section), we notice that it should be greater than the surface density of an
unperturbed standard accretion disc, because the squeezing process generally
increases $\Sigma$ \citep[see e.g. Figure 3 of][]{armitage02}.  This lower
limit is
\begin{equation} \Sigma_0 \simeq 1.6 \times 10^{5}
~\alpha^{-4/5}\dot{m}^{3/5}M_7^{1/5}r_2^{-3/5}~{\rm g~cm^{-2}}
\label{eq:sigma0}, \end{equation}
\citep[adapted from][]{kocsis12}, where $r_2=r/(10^2 R_S)$, $M_7 = M_p/(10^7
M_\odot)$ and $\alpha$ is the standard viscosity parameter \citep[][also see
Section~\ref{sec:numerical} for more details]{shakura73}.  Combining this
$\Sigma_0$ and the scale height $h_0$ derived earlier, we find a lower limit
\begin{equation} t_{\rm cool,0} = \kappa \Sigma_0 h_0 / (2 c) \simeq 0.25
~\alpha^{-4/5}\dot{m}^{8/5}M_7^{6/5}r_2^{-3/5}~{\rm yr},  \label{eq:uqt}
\end{equation}
for the cooling time-scale.

From Equation (\ref{eq:hrtt}) and the fact that
$t_{\rm cool} \geq t_{\rm cool,0}$, we derive 
\begin{equation}
h/r \geq 0.025~\alpha^{-  2 / 5} \dot{m}^{4 / 5} M_7^{1 / 10} r_2^{- 3 / 10} a_2^{- 2}[q(1 + q)]^{1 / 2},
\label{eq:h2r}
\end{equation}
where $a_2 = a / (10^2 R_S)$. Equation (\ref{eq:h2r}) can be simplified further:
Earlier works showed that the inner disc truncates at a radius 
of $r_{\rm in}\simeq n^{- 2 / 3} a$ due to the tidal effect,  where $n \ge 2$ is an integer determined by the strongest resonance
\citep{artymowicz94,liu03}. Replacing $r$ in Equation (\ref{eq:h2r})
with $r_{\rm in}$, we derive, for the outer boundary of the
inner disc,
\begin{equation}
h / r_{\rm in} > 0.025 ~\alpha^{- 2 / 5} \dot{m}^{4 / 5} M_7^{1 / 10} a_2^{- 23 / 10} n^{1 / 5}[q(1 + q)]^{1 / 2}.
\label{eq:h2r2}
\end{equation}

By equating the right-hand-side (RHS) of Equation~(\ref{eq:h2r2}) to $1$, we find
a critical separation $a_{\rm cri}$ for the SMBHB,
\begin{equation}
a_{\rm cri}\simeq 20~R_S~\alpha^{- 4 / 23} \dot{m}^{8 / 23} M_7^{1 / 23} n^{2 / 23}[q(1 + q)]^{5 / 23}.
\label{eq:ars}
\end{equation}
When $a$ reduces to about $a_{\rm cri}$, the disc aspect ratio will exceed
unity, first at $r\sim r_{\rm in}$. It is interesting that $a_{\rm cri}$
depends only weakly on each of the model parameters.

\section{Numerical simulation}
\label{sec:numerical}

We now use a one-dimensional numerical simulation to derive more accurately
$\Sigma$ and $h$ during the squeezing phase, taking into account both tidal and
viscous dissipation in the energy equilibrium.

Following \citet{armitage02}, we shrink the binary separation according to 
Equation~(\ref{eq:tgw}) and evolve the surface density of the inner
(circum-primary) disc by numerically integrating
\begin{equation}
\partial\Sigma/\partial t=-r^{-1}\partial\left(\Sigma r v_r\right)/\partial r,\label{eq:sigma}
\end{equation}
where
\begin{equation}
\Sigma r v_r=-3 r^{1/2}{\partial}\left(\nu\Sigma r^{1/2}\right)/{\partial r}
+2{\Sigma\Lambda}/{\Omega}\label{eq:cont}
\end{equation}
is the mass advection rate and $\Lambda$ the injection rate of specific angular
momentum due to the tidal torque of the binary \citep{lin86a}.  
To calculate $\Lambda$, we use
\begin{equation}
\Lambda=-0.5fq\Omega^{2}r^{2}\left(r/\Delta\right)^{4} \label{eq:lamb}
\end{equation}
\citep{armitage02}, with $f=10^{-2}$ a dimensionless parameter that constrains
the strength of the torque, $\Delta = \max \{R_h,h,|r-a|\}$ and $R_h =
a(q/3)^{1/3}$ is the Hills radius of the secondary black hole. We note that the
above scheme allows us to compute $v_r$ numerically without assuming the
relationship $v_r=-r/t_{\rm GW}$. We also implemented a smoothing of the tidal
torque \citep{tazzari15} but found little difference in the results.

The difference between our approach and the one from \citet{armitage02} lies in
the calculation of $h$. While in that work the authors did not allow $h$ to
vary with time, we evolve $h$ according to the so-called ``$\beta$-disc model''
\citep{shakura73}, also used by \citet{lodato09} and \citet{chang10}. In this
model, the viscosity $\nu$ is proportional to $h$ as well as the gas pressure,
which is only a fraction $\beta=p_{\rm gas}/(p_{\rm gas}+p_{\rm rad})$ of the
total pressure. As a result, $\nu=\alpha c_s h \beta$, where the sound speed
$c_s$ is defined as $c_s^2\equiv(p_{\rm gas}+p_{\rm rad})/\rho$ and $\rho =
\Sigma/(2h)$ is the volume density. To simplify the calculations of $p_{\rm
gas}$ and $p_{\rm rad}$, we assume that both quantities are determined by the
mid-plane temperature $T_c$ of the disc, such that $p_{\rm gas}=\rho k T_c /
(\mu m_p)$ and $p_{\rm rad} = 4\sigma T_c^4/(3c)$, with $k$ the Boltzmann
constant, $\sigma$ the Stefan-Boltzmann constant, $m_p$ the proton mass and
$\mu=0.615$ the mean particle mass in unit of $m_p$ for a plasma of solar
metallicity.

The computation of $h$ relies on three conventional assumptions which are also
valid in the system of our interest.  (i) Hydrostatic equilibrium in the
vertical direction, i.e. $c_s=\Omega h$. (ii) Heat is dissipated locally in the
form of radiation, i.e. $F=D_\nu+D_\Lambda$ with $D_\nu$ and $D_\Lambda$ as
described in Section~\ref{sec:heat}. (iii) Photons in the mid-plane are
transported to the disc surface by diffusion, so $F=4\sigma T_c^4/(3\tau)$.
These assumptions give us a system of three equations:
\begin{equation} 
\begin{aligned}
T_c &= \left[3\kappa \Sigma^2
\Omega\left(9\alpha c_s^2\beta-4
\Lambda\right)/(64\sigma)\right]^{1/4},\\
\beta &=
\left[1+8\sigma\mu m_p T_c^3 c_s/(3ck\Sigma\Omega) \right]^{-1},\\
c_s &= 8\sigma T_c^4/[3c\Omega\Sigma(1-\beta)].
\end{aligned}\label{eq:solution}
\end{equation}
Using these, as well as the surface density computed from the partial
differential equation, we solve $T_c$, $\beta$ and $c_s$ at each radius and
time step. More specifically, we derive a quartic function of $T_c$ and find
the one real and positive solution for our system. We then use this solution to
derive $\beta$, $c_s$, and finally $h$.

The simulation starts with a SMBHB of $M_p=10^7~M_\odot$ and $q=0.1$ at a
separation $a=100~R_S$, where the squeezing phase is expected to start
\citep{armitage02}. We set up an accretion disc around the primary black hole
using the surface density from Equation (\ref{eq:sigma0}) with $\dot{m}=0.01$.
A disc with a lower accretion rate would become radiatively inefficient and
already be thick \citep{narayan95}. An initial $\dot{m}$ higher than ours will
result in a higher $h$, as can be seen in Equation~(\ref{eq:h2r}). For this
reason, this simulation provides a lower limit on the scale height. The initial
disc is truncated to mimic the tidal interactions \citep{armitage02}, and we
consider a zero-torque inner boundary.

\begin{figure} \centerline{\includegraphics[width=\columnwidth]{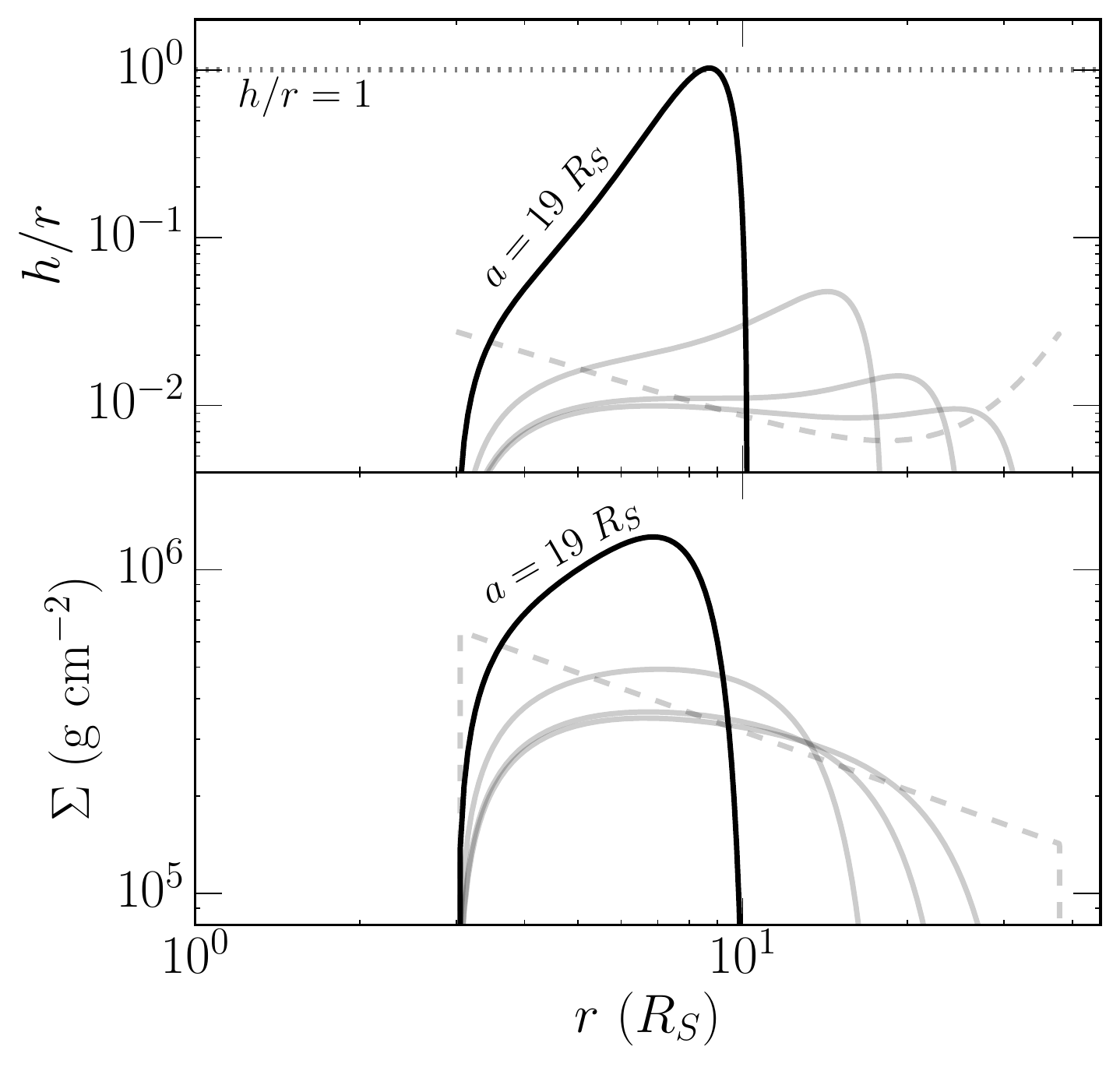}}
\caption{ Evolution of the disc aspect ratio (upper panel) and surface density
(lower panel) during the squeezing phase.  The gray dashed line in each panel
represents the initial condition. The black solid lines show the disc structure
in the final snapshot of our simulation, when the condition
$h/r=1$  (dotted line in the upper panel) is first met.  At this
point the binary separation is about $19~R_S$.  The gray solid
lines show the disc structure in three intermediate snapshots, with the disc
size shrinking with time.} \label{f:hr} \end{figure}

Figure \ref{f:hr} shows the evolution of $\Sigma$ and $h/r$.  Both quantities,
in general, increase over time.  We checked the aspect ratio $h/r$ in each time
step and stopped our simulation when the condition $h/r=1$ is met. This happens
at $a\simeq19~R_S$. By this point, the accretion rate, calculated as
$\dot{M}=2\pi r |v_r| \Sigma$ and Equation (\ref{eq:cont}), has increased to
almost  $\dot{M}_{\rm Edd}$ in most  of the disc. 

The numerical result agrees remarkably with our analytical prediction: If we
calculate $a_{\rm cri}$ using Equation~(\ref{eq:ars}) assuming $\dot{m}=1$ and
$n=2$, we find that $a_{\rm cri} \simeq 20~R_S$.  Moreover, throughout our
numerical simulation $\beta$ is much smaller than one, justifying the
assumption of a radiation-supported disc, as is adopted in
Sections~\ref{sec:heat} and \ref{sec:analytical}. We also find that only near
the end of the simulation does the cooling time-scale become comparable to the
GW-radiation time-scale.  Therefore, our assumption of local energy dissipation
is valid.

\section{Disc emission}\label{sec:rad}

\begin{figure} \centerline{\includegraphics[width=\columnwidth]{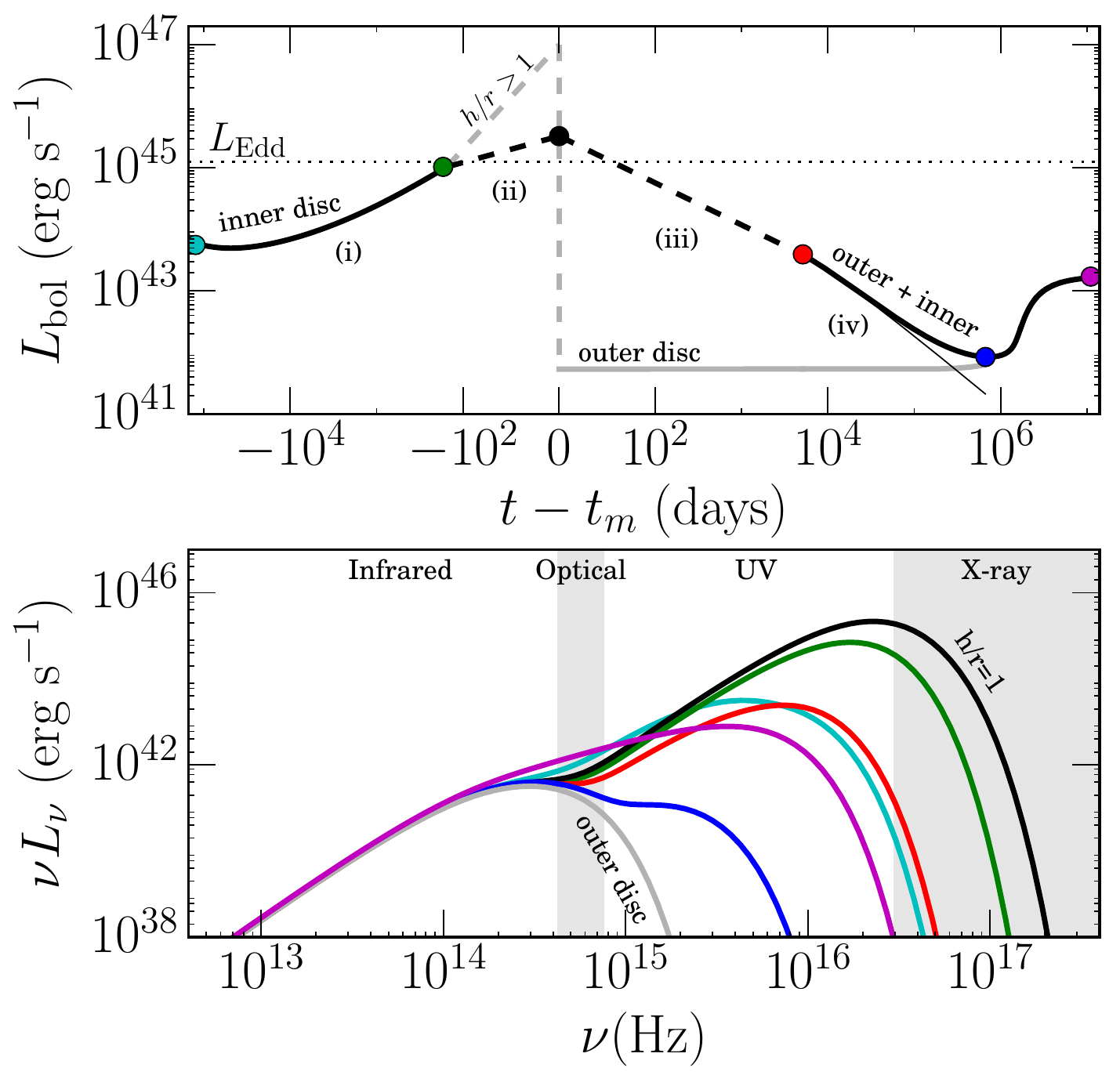}}
\caption{Upper panel: Evolution of the bolometric luminosity of the accretion
disc during four consecutive phases, namely (i) squeezing, (ii) decoupling,
(iii) cooling and (iv) recovering. The six colored dots mark six critical time
steps (see text). The black curves are our results and the
grey ones refer to the previous studies in comparison.  The solid lines (both
black and grey) are derived from our numerical simulations, and the dashed ones
are analytical results. The horizontal dotted line shows the Eddington
luminosity. 
Lower panel: SED at the six
critical times. The colors are the same as those of the dots in the top panel.
The four vertical bands in the background correspond to four EM wavebands, with
their names labelled on top.}\label{f:lsed} \end{figure}

The thickening of the inner disc will impact the EM counterpart in several
ways.  We divide the following discussion into four parts because the disc goes
through four consecutive phases where the dominant source powering the
radiation changes.  For each phase, we first identify the dominant power
source, then describe our method of calculating the radiation, and finally
discuss the results. The results are shown in Fig.~\ref{f:lsed}.  The top panel
shows the bolometric light curve, while the bottom panel shows the expected
spectral energy distribution (SED) at selected times.

(i) Squeezing phase: This is the phase that we modelled in the last two
sections.  During it, tidal heating dominates.  The simulation gives us the
surface temperature $T$, and since the disc is optically thick ($\tau\gg1$), we
can calculate the SED of each disc annulus using a black-body model
\citep{frank02}.  The resulting bolometric luminosity $L_{\rm bol}$ is shown as
a function of time in the top panel of Figure~\ref{f:lsed}, as the black solid
line labeled `i'.  The cyan and the blue dots, respectively, refer to the
initial condition and the last snapshot of the above numerical simulation.  We
can see that due to tidal heating the luminosity increases, until it reaches
the Eddington luminosity $L_{\rm Edd}=0.1\dot{M}_{\rm Edd}c^2$. 

(ii) Decoupling phase: When the disc becomes thick, the squeezing process is
ineffective ($D_\Lambda\approx 0$) because fluid elements of the inner disc can
cross the binary orbit through the horseshoe orbits or from outside the
equatorial plane, due to large effective viscosity and pressure gradient
\citep{lin79,papaloizou84,kocsis12,baruteau12}. The SMBHB goes through a second
decoupling--this time from the inner disc. Our code is unable to capture the
three-dimensional behaviour of the disc. However, we know that the material
ending up outside the binary orbit should remain hot and thick till the end of
the merger, because the cooling time-scale is longer than the GW-radiation
time-scale (Section~\ref{sec:analytical}). Moreover, this material should
maintain the same radial distance as that shown in Figure~\ref{f:hr}, because
tidal evolution is no longer important.  For these reasons, we assume that at
the time of the merger, $t=t_m$, the remnant inner disc has an aspect ratio of
$h/r=1$ and an outer boundary at $10~R_S$, the same as in the last snapshot of
our simulation.  Now using Equation~(\ref{eq:F}) we can calculate $F$ and
derive $L_{\rm bol}$, even though we do not know the exact surface density at
this point. The result is shown in the top panel of Figure~\ref{f:lsed} as the
black dot.  The black dashed line connecting the end of the squeezing phase and
the time of merger (with the label `ii') is a rough estimate of the luminosity
of the decoupling disc. It increases to a level above three times the Eddington
limit.

(iii) Cooling phase: Immediately after the merger, the only heating source is
viscosity, which is comparable to $D_\nu$ in the last evolutionary phase.  The
cooling rate, on the other hand, does not significantly change before and after
the merger because the temperature is similar.  Therefore, $F\simeq D_\Lambda$.
Then we have a situation in which cooling is more important than heating.  This
imbalance will lead to a drop of the temperature and a decrease of the disc
scale height, on a time-scale of $\left<t_{\rm cool}\right>$, the average
cooling time of the disc at the moment of the merger.  By the time
$t_m+\left<t_{\rm cool}\right>$, the disc would have cooled down such that $F$
becomes comparable to $D_\nu$ again. Considering $F=D_\nu$ and a constant scale
height, with the same mass and size than the last snapshot of our simulation,
we can apply the standard-disc model and obtain $h,\,T,\,\Sigma$ as a function
of $r$.  Those assumptions are motivated by noticing that the total luminosity
never significantly exceeds the Eddington limit.  Consequently, the mass loss
due to disc wind, which relies on a super-Eddington luminosity
\citep[e.g.][]{lodato09,tazzari15}, would be much weaker than previously has
been thought.

The luminosity at this time is shown as the red dot in the top panel of
Figure~\ref{f:lsed}. The evolution of the bolometric luminosity during the
cooling phase is represented by the black dashed line connecting the black and
the red dots (with the label `iii').  We can see that it drops by more than one
order of magnitude, indicating that the disc cools down significantly.

(iv) Recovering phase: After the condition $F=D_\nu$ is re-established, the
following evolution of the disc is dominated by viscosity. We simulate it with
our one-dimensional code setting $\Lambda=0$. The initial condition is the same
as the solution derived in the previous cooling phase.  Since the viscous
time-scale of this new inner disc is comparable to that of the original outer
disc (at $r>10^2~R_S$), we no longer can assume that the outer disc is
invariant. Therefore, we include the outer disc in the simulation, with an
initial condition given by Equation \ref{eq:sigma0} and $\dot{m}=0.01$, as well
as a constant outer boundary. The resulting luminosity is shown in the top
panel of Figure~\ref{f:lsed} as the black solid line to the RHS of the red dot
(with the label `iv').  It first decreases, because the effective accretion
rate of the inner disc drops as it expands. By the time the inner and the outer
discs overlap (blue dot), the luminosity has reached a minimum. Afterwards,
fresh material from the outer part of the disc refills the inner disc, driving
the effective accretion rate towards the equilibrium, i.e.  $\dot{m}=0.01$. As
a result, the entire disc returns to the thin-disc solution and the luminosity
recovers the original value of $0.01L_{\rm Edd}$ (purple dot).

Having understood the light curve of a thick disc, let us now compare it with
that of a thin disc, which is presented in the top panel of Figure~\ref{f:lsed}
as the grey curves. Three important differences appear.

First, the luminosity of the thin disc rises sharply during the last ${\cal
O}(10^2)$ days of the merger, from $L_{\rm Edd}$ to more than $10$ times
higher.  It has been proposed that this ``precursor'' could be used to alert GW
detectors for follow ups \citep{chang10}.  We now see that such a large
enhancement is unphysical because the power source--tidal heating--would
already have shut down. The luminosity of the thick disc, according to our
calculation, increases only $2-3$ times during this period.

Second, the luminosity of the thin disc drops immediately after the merger by
about four orders of magnitude. This behavior is caused by a complete depletion
of the inner disc by the squeezing mechanism, leaving only the emission of the
outer disc. It has been pointed out that this cut-off and the later recovering
of the luminosity (due to a refilling of the inner cavity) can be used to
identify black-hole mergers \citep{milosavljevi05}.  Besides, it also has been
suggested that the disappearance of the inner disc could explain the
interruption of jet activity seen in a sample of radio galaxies
\citep{liu03,liu04}. Both proposals would have difficulties in the light of our
new results, because the inner disc never completely disappears.

The third difference is related to the second one, but is more clearly seen in
the evolution of the SED, which is shown in the lower panel of
Figure~\ref{f:lsed}.  If the inner disc is completely depleted, the SED after
the merge would come entirely from the outer disc (grey solid curve) where the
corresponding UV and X-ray luminosities are negligible. This part of the SED
re-brightens on a time-scale of $10^6$ days because the outer disc refills the
inner cavity on the viscous time-scale.  In our model, however, the UV/X-ray
emission remains present even after the merger (black SED) because the inner
disc does not disappear.  Only on a time-scale of ${\cal O}(10^4)$ days after
the merger does the UV/X-ray radiation fade away because of cooling (compare
the black and red SEDs). This behavior is opposite to that suggested by the
thin-disc models.

Therefore, we have seen that the thickening and the second decoupling of the
inner disc has important implications for the detectability of the EM
counterparts.  It is worth noting that the residual inner disc remains
gravitationally bound to the merged SMBH even though the black hole will
receive a recoil velocity due to anisotropic GW radiation \citep{centrella10}.
During this recoil the disc could be shock-heated to an even higher
temperature, because of the shear induced by a passing GW \citep{kocsis08}, a
loss of gravitational mass \citep{milosavljevi05,schnittman08,megevand09}
and/or the orbital change relative to the recoiling SMBH
\citep{lippai08,shields08,schnittman08}.  Existence of such a hot disc also
opens many possibilities of detecting recoiling SMBHs. 

\section*{Acknowledgments}

This work is supported by the China-CONICYT fellowship (No. CAS15002), by
CONICYT through FONDECYT (1141175) and Basal (PFB0609) grants, and partly by
the Strategic Priority Research Program ``Multi-wavelength gravitational wave
universe'' of the Chinese Academy of Sciences (No.  XDB23040100).  XC thanks
for the hospitality and support of the Gravitational Wave Astrophysics Group at
the National Astronomical Observatories of China.

\bibliographystyle{mnras}

\bsp
\label{lastpage}
\end{document}